\documentclass[11pt]{article}
 \usepackage[top=1in,bottom=1in,left=1.5in,right=1.5in]{geometry}
  \usepackage{amsmath,amssymb}
  \usepackage{epsfig}
  \usepackage{tikz}
  \usepackage{latexsym}

  \newcommand{\proof}{\noindent \textbf{Proof.\ }}

  \newcommand{\qed}{\hspace*{\fill} $\Box$\\}

  \numberwithin{equation}{section} 
\newtheorem{theorem}{Theorem}

\newtheorem{corollary}[theorem]{Corollary}

\newtheorem{remark}[theorem]{Remark}

\begin{document}

 \title{\textbf{Some Results on Dominating\\ Induced Matchings}}

\author{
    S. Akbari$^1$ \footnote{E-mail addresses: s\_akbari@sharif.edu (S. Akbari), hossein.bktsh@gmail.com (H. Baktash), amin.bahjati@gmail.com (A. Behjati), behmaram@tabrizu.ac.ir (A. Behmaram), mohammadroghani43@gmail.com (M. Roghani)}, H. Baktash$^2$, A. Behjati$^2$, A. Behmaram$^3$,  M. Roghani$^2$
}

\date{\small{
    $^1$Department of Mathematical Sciences, Sharif University of Technology, Tehran, Iran\\
    $^2$Department of Computer Engineering, Sharif University of Technology, Tehran, Iran.\\
    $^3$Faculty of Mathematical Sciences, University of Tabriz, Tabriz, Iran
}}

\maketitle

\begin{abstract}

 Let $G$ be a graph, a dominating induced matching (DIM) of $G$ is an induced matching that dominates every edge of $G$. In this paper we show that if a graph $G$ has a DIM, then $\chi(G) \leqslant  3$. Also, it is shown that if $G$ is a connected graph whose all edges can be partitioned into DIM, then $G$ is either a regular graph or a biregular graph and indeed we characterize all graphs whose edge set can be partitioned into DIM. Also, we prove that if $G$ is an $r$-regular graph of order $n$ whose edges can be partitioned into DIM, then $n$ is divisible by $\binom{2r - 1}{r - 1}$ and $n = \binom{2r - 1}{r - 1}$ if and only if $G$ is the Kneser graph with parameters $r-1$, $2r-1$.

\end{abstract}



\section{Introduction}
Let $G$ be a simple graph with the vertex set $V(G)$ and the edge set $E(G)$.
As usual $|V(G)| = n$ and $|E(G)| = m$ denote the number of vertices and edges of $G$. The minimum and the maximum degree of a graph $G$ are denoted by $\delta(G)$ and $\Delta(G)$, respectively. For a vertex $v$, $N(v)$ denotes the set of neighbors of $v$ in $G$. A cycle of order $r$ is denoted by $C_r$. Also, a \textit{vertex $k$-coloring} of a simple graph $G$ is defined as an assignment of one element from a set $C$ of $k$ colors to each vertex in $V(G)$ such that no two adjacent vertices have the same color, and the minimum $k$ for which $k$-coloring is possible, is $\chi(G)$. A \textit{Kneser graph} $KG_{n,k}$ is the graph whose vertices correspond to the $k$-subsets of a set of $n$ elements, and two vertices are adjacent if and only if their two corresponding sets are disjoint. The Petersen graph is an example of Kneser graph $KG_{5,2}$ given in Figure 1. A $BG_{m,n}$ is a bipartite graph, whose vertices in one part correspond to the $m$-subsets of a set of $m+n-1$ elements and vertices in the other part correspond to $n$-subsets of the same set and two vertices in different parts are adjacent if and only if they are disjoint. A \textit{biregular} graph is a bipartite graph $G = (X, Y)$ in which any two vertices in $X$ have the same degree and any two vertices in $Y$ have the same degree.

\begin{figure}[!h]
\centering
\begin{tikzpicture}[every node/.style={font=\fontsize{6}{0}\selectfont}, scale=0.8]
\draw[fill=black] (1.9021, 0.6180) circle (2pt);
\draw[fill=black] (0, 2) circle (2pt);
\draw[fill=black] (-1.9021, 0.6180) circle (2pt);
\draw[fill=black] (-1.1755, -1.6180) circle (2pt);
\draw[fill=black] (1.1755, -1.6180) circle (2pt);

\draw[fill=black] (0.9510, 0.3090) circle (2pt);
\draw[fill=black] (0, 1) circle (2pt);
\draw[fill=black] (-0.9510, 0.3090) circle (2pt);
\draw[fill=black] (-0.5877, -0.8090) circle (2pt);
\draw[fill=black] (0.5877, -0.8090) circle (2pt);
\node at (0, 2.3) {\{1, 2\}};
\node at (-2.5521, 0.6180) {\{4, 5\}};
\node at (-1.4755, -1.9880) {\{2, 3\}};
\node at (1.4755, -1.9880) {\{1, 5\}};
\node at (2.5521, 0.6180) {\{3, 4\}};

\node at (1.1610, -0.0190) {\{2, 5\}};
\node at (0.55, 1.05) {\{3, 5\}};
\node at (-0.8510, 0.5990) {\{1, 3\}};
\node at (-0.9577, -0.5390) {\{1, 4\}};
\node at (0.2777, -1.1090) {\{2, 4\}};
\draw[thick] (1.9021, 0.6180) -- (0, 2) -- (-1.9021, 0.6180) -- (-1.1755, -1.6180) -- (1.1755, -1.6180) -- (1.9021, 0.6180);
\draw[thick] (0.9510,0.3090) -- (-0.9510,0.3090) -- (0.5877,-0.8090) -- (0,1) -- (-0.5877,-0.8090) -- (0.9510,0.3090);
\draw[thick] (0, 2) -- (0,1);
\draw[thick] (-1.9021, 0.6180) -- (-0.9510,0.3090);
\draw[thick] (-1.1755, -1.6180) -- (-0.5877,-0.8090);
\draw[thick] (1.1755, -1.6180) -- (0.5877,-0.8090);
\draw[thick] (1.9021, 0.6180) -- (0.9510,0.3090);
\end{tikzpicture}
\caption{\protect\centering The Petersen graph is isomorphic to the Kneser graph $KG_{5,2}$.}
\label{fig:petersen}
\end{figure}
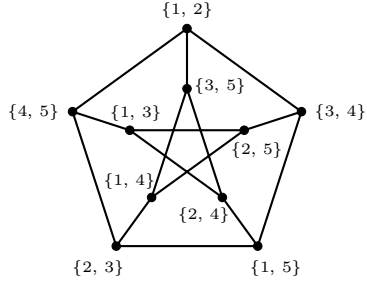

A \textit{matching} of $G$ is a set of mutually non-adjacent edges of $G$. An \textit{induced matching} (IM) is a matching having no two edges joined by an edge. In other words, $M$ is an induced matching (also called \textit{strong matching}) of $G$ if the subgraph of $G$ induced by $V(M)$ is 1-regular. A maximum induced matching is an induced matching of maximum cardinality. Finding maximum IM is NP-hard \cite{Cameron}. The concept of domination in graphs appears as a natural model for facility location problems and has many applications in design and analysis of communication networks, network routing, and coding theory, see \cite{brand1} and \cite{brand2}.

\vspace{2mm}
An edge of $G$ dominates itself and every edge adjacent to it. A \textit{dominating induced matching} (DIM) (or \textit{efficient edge dominating set} (EEDS) in some papers) of $G$ is an induced matching that dominates every edge of $G$. We denote $dim(G)$, the size of the smallest DIM in $G$. Clearly, not every graph has DIM, for instance consider $C_4$. Denote $E(u)$ as the set of all edges incident with $u$. For every edge $e$, let $D_e$ be the set of edges that are dominated by $e$. Note that for a DIM, all edges are being dominated by exactly one of the edges in DIM. The cycle $C_{n}$ has an DIM if and only if $n = 3k$. Dominating induced matchings have been extensively studied by several authors, for instance see \cite{brandst}, \cite{cardoso}, \cite{hertz}, and \cite{cardozo}.

\section{Existence and Some Properties of \\Dominating Induced Matching}

In this section, we prove that if a graph $G$ has a DIM, then its chromatic number does not exceed 3. Furthermore, we show that if $G$ has a DIM, then the size of all DIM of $G$ are the same. First, we start with the following result.

\begin{theorem}

If a graph $G$ has a \textup{DIM}, then $\chi(G) \leqslant  3$.

\end{theorem}

\proof Let $M= \{v_{1} w_{1}, \ldots, v_{k}w_{k} \}$ be a DIM for the graph $G$. For $ i = 1, \ldots, k$ we color $v_{i}$ and $w_{i}$ by colors $1, 2$, respectively and color the remaining vertices by
color $3$. By the definition of DIM, this coloring is a proper vertex coloring by $3$ colors and so $ \chi(G) \leqslant  3$.\qed \\

\noindent
In the following result we provide an upper bound for the size of a graph which has a DIM.
\begin{theorem}

If a graph $G$ of order $n$ has a \textup{DIM}, then $|E(G)| \leq \frac{n^2 + n}{4}$.

\end{theorem}

\proof Let $D$ be a DIM of $G$. For every vertex $v \in V(D)$, obviously, there are exactly $dim(G)$ edges whose two ends are in $V(D)$. Moreover, since $V(G) \setminus V(D)$ is an independent set, there are no edges whose ends are in $V(G) \setminus V(D)$. Also, the number of edges between $V(D)$ and $V(G) \setminus V(D)$ is at most $2dim(G)(n - 2dim(G))$. Therefore, we have the following inequality:
$$|E(G)| \leq 2dim(G)(n - 2dim(G)) + dim(G).$$
This inequality will be maximized whenever $dim(G) = \frac{n}{4}$. Therefore the following holds:
$$|E(G)| \leq \frac{n}{2}(n - \frac{n}{2}) + \frac{n}{4} = \frac{n^2 + n}{4} \text{ ,}$$\\
as desired.\qed

\noindent
The next result shows that the size of all DIM of $G$ are the same.

\begin{theorem}
Let $G$ be a graph. If $D_1$ and $D_2$ are two \textup{DIM} of $G$, Then $|D_1| = |D_2|$.
\end{theorem}
\proof
Obviously, for every $e \in E(D_1), |D_e \cap D_2| \leq 1$, because $e$ cannot be dominated by more than one edge in $D_2$. Since $E(G) = \bigcup_{e \in E(D_1)}^{} D_e$, one can see that $|D_2| \leq |D_1|$. Similarly, we have $|D_1| \leq |D_2|$ and the proof is complete.\qed

\begin{theorem}
Let $G$ be a graph of order $n$ with at least one \textup{DIM} such that $\delta(G) \geq 2$. Then the following inequalities hold:
$$ \frac{\delta(G)}{\Delta(G)-1} \leq  \frac{2dim(G)}{n - 2dim(G)} \leq  \frac{\Delta(G)}{\delta(G)-1}$$
In particular, if $G$ is a $k$-regular graph, then
$dim(G)=\frac{nk}{4k-2}.$
\end{theorem}

\proof Let $D$ be a DIM of $G$. For every vertex $v \in V(D)$, since $v$ is adjacent to exactly one edge of $D$, $v$ has at least $\delta(G) - 1$ and at most $\Delta(G) - 1$ neighbors out of $V(D)$. Also for every vertex $v \notin V(D)$, since $V(G) \setminus V(D)$ is an independent set, $v$ has at least $\delta(G)$ and at most $\Delta(G)$ neighbors in $V(D)$. Then by double counting of the number of edges between $V(D)$ and $V(G) \setminus V(D)$, we find,

$$\delta(G)(n-2dim(G)) \leq 2dim(G)(\Delta(G)-1)$$
$$\Delta(G)(n-2dim(G)) \geq 2dim(G)(\delta(G)-1).$$
\\
From these two inequalities, the proof of the first part is complete. Now, if $G$ is a $k$-regular graph, then the following hold:
$$ \frac{k}{k-1} \leq \frac{2dim(G)}{n - 2dim(G)} \leq \frac{k}{k-1}.$$
This implies that,
$$ \frac{k}{k-1} = \frac{2dim(G)}{n - 2dim(G)},$$
and so we have,
$$dim(G)=\frac{nk}{4k-2}.$$\qed

\noindent
Now, we have an immediate corollary.

\begin{corollary}
Let $G$ be a $k$-regular graph of order $n$ with at least one \textup{DIM}. Then, $4k - 2 | nk$.
\end{corollary}

\noindent
In the following, we study the size of intersection of a DIM with cycles in a graph.
\begin{theorem}

Let $G$ be a graph. Then for every cycle $C_r$ of $G$, and each \textup{DIM}, $D$ of $G$ the following hold:
$$|E(C_r) \cap E(D)| \leq \frac{r}{3}, \quad\quad |E(C_r) \cap E(D)| \equiv r\pmod{2}.$$
\end{theorem}

\proof
Let $e \in E(D)$. If $e \in E(C_r)$, then $e$ dominates exactly $3$ edges of $C_r$, and if $e \notin E(C_r)$, then $e$ dominates an even number of edges in $C_r$. Since every edge of $C_r$ is dominated by exactly one edge of $D$, then $|E(C_r) \cap E(D)| \leq \frac{r}{3}$ and the parity of $r$ equals to the parity of $|E(C_r) \cap E(D)|$.\qed \\

\noindent
Now, we have the following corollary.

\begin{corollary}
Let G be a graph with a \textup{DIM}, D. Then the following hold:
\renewcommand\labelenumi{(\roman{enumi})}
\renewcommand\theenumi\labelenumi
\begin{enumerate}
    \item Each $C_3$, $C_5$ or $C_7$ in $G$, meets $D$ in exactly one edge.
    \item $D$ contains no edge of $C_4$ in $G$.
\end{enumerate}
\end{corollary}

\section{Edge Partitioning of Graphs into \\Dominating Induced Matching}

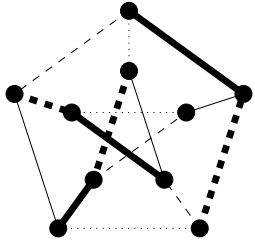
\begin{figure}[!h]
\centering
\begin{tikzpicture}[scale=0.8]
\draw[fill=black] (1.9021, 0.6180) circle (4pt);
\draw[fill=black] (0, 2) circle (4pt);
\draw[fill=black] (-1.9021, 0.6180) circle (4pt);
\draw[fill=black] (-1.1755, -1.6180) circle (4pt);
\draw[fill=black] (1.1755, -1.6180) circle (4pt);

\draw[fill=black] (0.9510, 0.3090) circle (4pt);
\draw[fill=black] (0, 1) circle (4pt);
\draw[fill=black] (-0.9510, 0.3090) circle (4pt);
\draw[fill=black] (-0.5877, -0.8090) circle (4pt);
\draw[fill=black] (0.5877, -0.8090) circle (4pt);
\draw[line width=0.9mm] (1.9021, 0.6180) -- (0, 2);
\draw[dashed] (0, 2) -- (-1.9021, 0.6180);
\draw[line width=0.1mm] (-1.9021, 0.6180) -- (-1.1755, -1.6180);
\draw[dotted] (-1.1755, -1.6180) -- (1.1755, -1.6180);
\draw[dashed, line width=0.9mm] (1.1755, -1.6180) -- (1.9021, 0.6180);

\draw[dotted] (0.9510,0.3090) -- (-0.9510,0.3090);
\draw[line width=0.9mm] (-0.9510,0.3090) -- (0.5877,-0.8090);
\draw[line width=0.1mm] (0.5877,-0.8090) -- (0,1);
\draw[dashed, line width=0.9mm] (0,1) -- (-0.5877,-0.8090);
\draw[dashed] (-0.5877,-0.8090) -- (0.9510,0.3090);

\draw[dotted] (0, 2) -- (0,1);
\draw[dashed, line width=0.9mm] (-1.9021, 0.6180) -- (-0.9510,0.3090);
\draw[line width=0.9mm] (-1.1755, -1.6180) -- (-0.5877,-0.8090);
\draw[dashed] (1.1755, -1.6180) -- (0.5877,-0.8090);
\draw[line width=0.1mm] (1.9021, 0.6180) -- (0.9510,0.3090);
\end{tikzpicture}
\caption{Partitioning of all edges of Petersen graph into five DIM.}
\label{fig:petersen_part}
\end{figure}

In this section, we show that the edge set of some families of Kneser graphs can be partitioned into dominating induced matchings. Also, we show that if $G$ is a connected graph whose all edges can be partitioned into DIM, then $G$ is either a regular graph or a biregular graph. For example, all edges of the Petersen graph can be partitioned into five DIM, as shown in Figure \ref{fig:petersen_part}.

\begin{theorem}\label{partition results}
Let $G$ be a connected graph whose all edges can be partitioned into \textup{DIM}. Then $G$ is either a regular graph or a biregular graph. Moreover, the number of \textup{DIM} is $d(u) + d(v) - 1$, for each $uv \in E(G)$.
\end{theorem}
\proof Let $u$ and $v$ be two adjacent vertices of $G$. Note that if one of the edges of $E(u) \cup E(v)$ is contained in a DIM, then no other edge of $E(u) \cup E(v)$ is in the DIM. Also, every DIM has at least one edge in $E(u) \cup E(v)$. Thus, every DIM has exactly one edge in $E(u) \cup E(v)$. Hence, if $uv \in E(G)$, then the number of DIM that partition the edges of $G$ is $d(u) + d(v) - 1$. Choose a vertex $w$. By earlier statement, we see that the degree of vertices in $N(w)$ are the same. With the same argument, the degree of all vertices of distance $2$ from $w$ is $d(w)$, and so on. Hence $G$ is either a regular graph, when the degree of $w$ and its neighbors are the same, or a biregular graph, when the degree of $w$ and its neighbors are not the same.\qed

\noindent
Before characterization of all graphs whose edges can be partitioned into DIM, we need two following results.

\begin{theorem}
Let $r$ be a positive integer. Then all edges of $KG_{2r-1, r-1}$ can be partitioned into $2r-1$ \textup{DIM}.
\end{theorem}

\proof First we define an edge coloring for $KG_{2r-1, r-1}$ with the colors $1,\ldots, 2r-1$. Let $XY$ be an edge of $KG_{2r-1, r-1}$, where $X$ and $Y$ are $(r-1)$-subsets of $S = \{1, \ldots, 2r-1\}$. We color the edge $XY$ by $S \setminus (X \cup Y)$ (Note that $S \setminus (X \cup Y)$ has one element). Assume that $C_1,\ldots, C_{2r-1}$ are all color classes. We claim that for $i = 1,\ldots, 2r-1$, $C_i$ is a DIM. Let $XY, XY' \in C_i$. Thus we have $Y = Y' = S \setminus (X \cup \{i\})$, a contradiction. Hence all edges of $C_i$ form a matching. Now, assume that $XY, X'Y' \in C_i$ and there exists the edge $X'Y$ between $XY$ and $X'Y'$, where $X'Y \in C_j$. Clearly, $i,j \notin X' \cup Y$. On the other hand since $X'\cap Y = \varnothing$, we have $|X' \cup Y| = 2r - 2$, contradiction. Let $Z$ and $Z'$ be two vertices of $KG_{2r-1, r-1}$ and no edge in $C_i$ is incident with $Z$ and $Z'$. Obviously, $i \in Z \cap Z'$ and so $Z$ and $Z'$ are not adjacent. Therefore for $i = 1,\ldots, 2r - 1$, $C_i$ is a DIM. The proof is complete.\qed

\noindent
Now, we would like to prove that the edge set of $BG_{r-1,s-1}$ can be partitioned into DIM, where $r$ and $s$ are two positive integers.

\begin{theorem}
For every positive integers $r, s$, $E(BG_{r-1,s-1})$ can be partitioned into $r+s-1$ \textup{DIM}.
\end{theorem}

\proof First, we introduce an edge coloring for $BG_{r-1,s-1}$ using the colors $S = \{1,\ldots, r + s - 1\}$. For the edge $XY$, we color $XY$ by the color $S \setminus (X \cup Y)$. For $i = 1,\ldots, r + s - 1$ denote the $i$-th color class by $C_i$. Let $XY$ and $XY'$ be two edges in $C_i$. Then, $Y = Y' = S \setminus (X \cup \{i\})$, a contradiction. Now, Let $XY, X'Y' \in C_i$ and $XY' \in C_j$. Thus $i,j \notin X \cup Y'$ and since $|X \cup Y'| = r + s - 2$, we get a contradiction. Let $Z$ and $Z'$ be two vertices of $BG_{r-1,s-1}$ and no edge of $C_i$ is incident with $Z$ and $Z'$. Obviously, $i \in Z \cap Z'$ and so $Z$ and $Z'$ are not adjacent. Therefore, for $i = 1,\ldots, r+s-1$, $C_i$ is a DIM.\qed\\

\noindent
Now, we are in a position to prove our main theorem.

\begin{theorem} \label{main theorem}
Let $G$ be an $r$-regular graph and $E(G)$ can be partitioned into \textup{DIM}. Then there exists a list of assignment $L: V(G) \rightarrow A$, where $A$ is the set of all $(r-1)$-subsets of $S = \{1, \ldots, 2r - 1\}$, which has the following properties:
\renewcommand\labelenumi{(\roman{enumi})}
\renewcommand\theenumi\labelenumi
\begin{enumerate}
    \item For every $uv \in E(G)$, $L(u) \cap L(v) = \varnothing$,
    \item $L$ is surjective,
    \item For every $a, b \in A$, $|L^{-1}(a)| = |L^{-1}(b)|$.
\end{enumerate}
\end{theorem}

\proof By Theorem \ref{partition results}, let $M_1, M_2, \ldots, M_{2r-1}$ be all DIM which partition $E(G)$. Now, we define an edge coloring for $G$. For every $e \in E(G)$ define $c(e) = t$ if and only if $e \in M_t$. Now, we introduce the list assignment $L: V(G) \rightarrow A$ as follows. For every $v \in V(G)$, define $L(v) = S \setminus \{c(e)\,|\, \text{\textit{e} is incident with \textit{v}}\}$:
\renewcommand\labelenumi{(\roman{enumi})}
\renewcommand\theenumi\labelenumi
\begin{enumerate}
\item With no loss of generality, one can assume that $G$ is a connected graph. First we prove Part (i). By contradiction assume that $uv \in E(G)$ and $j \in L(u) \cap L(v)$. By the definition, no edge incident with $u$ or $v$ has color $j$. Hence $M_j$ does not dominate $uv$, a contradiction. So (i) is proved.
\item First we claim that for every $v \in V(G)$, if $a$ is an $(r-1)$-subset of $S$ such that $L(v) \cap a = \varnothing$, then there is a vertex $u \in N(v)$ such that $L(u) = a$. Suppose that $u, u' \in N(v)$. It is not hard to see that, $c(uv) \notin L(u)$ and $c(uv) \in L(u')$. Thus, $L(u) \neq L(u')$. Since $d(v) = r$ and $S \setminus L(v)$ have exactly $r$, $(r - 1)$-subsets, we conclude that every $(r-1)$-subset of $S \setminus L(v)$ appears as a list of a vertex in $N(v)$ once and the claim is proved. Let $v \in V(G)$ and $L(v) = a$. Assume that $b$ is an $(r-1)$-subset of $S$ and $k = |a \cap b|$. By induction on $k$ we show that there exists a vertex $u \in V(G)$ such that $L(u) = b$. For $k = 0$, there is nothing to prove. Let the assertion hold for $k-1$ and $k=|a \cap b|$. Assume that $x \in a \cap b$ and $y \in S \setminus (a \cup b)$. Let $c = (b \setminus \{x\}) \cup \{y\}$. Clearly, $|a \cap c| = k - 1$ and so by the induction hypothesis there exists $w \in V(G)$ such that $L(w) = c$. Suppose that $d = S \setminus (\{x\} \cup c)$. Since $L(w) \cap d = \varnothing$, By the claim, there exists $f \in V(G)$ which is adjacent to $w$ and $L(f) = d$. Since $b \cap d = \varnothing$, by the claim there exists $g \in V(G)$ such that $L(g) = b$.
\item Since the restriction of $L$ on each connected component of $G$ is surjective, it suffices we prove that if $a \cap b = \varnothing$, then $|L^{-1}(a)| = |L^{-1}(b)|$. By the claim of Part (ii) we note that for every $v \in L^{-1}(a)$ there exists $u \in L^{-1}(b)$ which is adjacent to $v$, such that $c(uv) = S \setminus (a \cup b)$. Since by the claim of Part (ii) for every $v' \in N(u), c(uv) \neq c(uv')$, then every vertex in $L^{-1}(a)$ is adjacent to exactly one vertex in $L^{-1}(b)$. This implies that $|L^{-1}(a)| \leq |L^{-1}(b)|$. Similarly, $|L^{-1}(b)| \leq |L^{-1}(a)|$ and the proof is complete.\qed
\end{enumerate}

\begin{remark}
\rm{By the same proof of Theorem \ref{main theorem} one can see that the three following statements hold for biregular graphs $G(X, Y)$ in which $E(G)$ can be partitioned into \rm{DIM}, where for every $u \in X$ and $v \in Y$, $d(u) = x$ and $d(v) = y$. If we replace $L$ (define in Theorem \ref{main theorem}) with $L' : V(G) \rightarrow A_1 \cup A_2$, where $A_1$ and $A_2$ are respectively the $(x - 1)$-subsets and $(y - 1)$-subsets of $S = \{1, 2, \ldots, x + y - 1\}$ and $L'(X) \subseteq A_1$, $L'(Y) \subseteq A_2$:
\renewcommand\labelenumi{(\roman{enumi})}
\renewcommand\theenumi\labelenumi
\begin{enumerate}
\item For every $uv \in E(G)$, $L'(u) \cap L'(v) = \varnothing$,
\item $L'$ is surjective,
\item For every $a, b \in A_1 \cup A_2$, $|L'^{-1}(a)| = |L'^{-1}(b)|$.
\end{enumerate}}
\end{remark}

\noindent
Now, we have the following corollary.

\begin{corollary}
Let $G$ be an $r$-regular graph and $E(G)$ can be partitioned into \textup{DIM}. Then $|V(G)|$ is divisible by $\binom{2r - 1}{r - 1}$ and $|V(G)| = \binom{2r - 1}{r - 1}$ if and only if $G$ is $KG_{2r-1, r-1}$.

\end{corollary}

\end{document}